\documentclass{article}
\usepackage{spconf,amsmath,graphicx,hyperref}
\usepackage{cite}
\usepackage{amssymb,amsfonts}
\usepackage{booktabs}
\usepackage{balance}
\usepackage{tikz}
\usetikzlibrary{arrows.meta}

\def\BibTeX{{\rm B\kern-.05em{\sc i\kern-.025em b}\kern-.08em
    T\kern-.1667em\lower.7ex\hbox{E}\kern-.125emX}}

\title{Soft Posterior Speaker Injection for Multi-Talker Speech Recognition}
\name{ \small Jian Zhu$^{1}$, Jun Sun$^{1}$, Jiang Yang$^{1,\ast}$, Ying Zhou$^{1,\ast}$, Cheng Luo$^{2}$, Yang Ai$^{3}$, Hong-Hao Sun$^{3}$, Junhui Shi$^{1}$, Li-Rong Dai$^{3}$
\thanks{$^{\ast}$Corresponding author.}}
\address{$^{1}$ Zhejiang Lab, Hangzhou, China\quad qijian.zhu@zhejianglab.com
\\	$^{2}$ Zhejiang International Studies University, Hangzhou, China\quad luo\_cheng@zisu.edu.cn
\\	$^{3}$ University of Science and Technology of China, Hefei, China\quad lrdai@ustc.edu.cn
}
\begin{document}
\ninept
\setlength{\textfloatsep}{6pt plus 1pt minus 2pt}
\setlength{\floatsep}{6pt plus 1pt minus 2pt}
\setlength{\intextsep}{6pt plus 1pt minus 2pt}
\setlength{\abovecaptionskip}{3pt}
\setlength{\belowcaptionskip}{0pt}
\maketitle
\begin{abstract}
Multi-talker automatic speech recognition (MT-ASR) remains challenging in the presence of overlapping speech. Hard segmentation introduces irreversible errors, whereas serialized output training (SOT) avoids explicit segmentation but does not condition a pretrained encoder on speaker activity. We propose Soft Posterior Speaker Injection (SPSI). A Soft Posterior Head predicts per-frame speaker posteriors $\hat{\mathbf{P}}$ and injects them into Whisper through Multi-layer Feature-wise Linear Modulation (MFLM) and Speaker Memory Prompts (SMP). The benefit of SPSI is largest where overlap is heaviest and under domain transfer. On controlled two-speaker LibriSpeech overlap, SPSI reduces concatenated minimum-permutation word error rate (cpWER) from $61.5\%$ to $60.0\%$ in the high-overlap bin, and from $51.9\%$ to $51.0\%$ on the full set, relative to SOT. By contrast, Speaker CE, SD-CTC, SA-DiCoW, and Pipeline (oracle/est.\ VAD) do not outperform SOT. Freeze-posterior overlap-heavy adaptation reduces held-out LibriCSS cpWER from $42.3\%$ to $36.8\%$ on sessions $8$--$9$, a $5.5$-point gain over SOT. The source code is available at \url{https://github.com/HackerHyper/SPSI}.
\end{abstract}

\section{Introduction}
\label{sec:intro}
Multi-talker automatic speech recognition (MT-ASR) aims to transcribe a
conversation into per-speaker text streams when speakers may
overlap~\cite{watanabe2020chime,chen2020continuous,seki2018end,kanda2022tsot,moriya2025aft}, as is common in meetings.
The recognizer must recover not only the words, but also who spoke them.
Pretrained encoder--decoder models such as Whisper~\cite{radford2023whisper}
are a practical backbone, yet they are trained mainly on single-talker audio.

Existing speaker-activity interfaces are either irreversible, absent, or too coarse.
Cascaded systems first cut the mixture with voice activity detection (VAD),
diarization, or source
separation~\cite{hershey2016deep,yu2017permutation,zmolikova2019speakerbeam}.
A boundary error of tens to hundreds of milliseconds cannot be undone:
onsets are truncated and competing speech leaks into the wrong stream.
Serialized output training (SOT)~\cite{kanda2020serialized,kanda2022tsot}
avoids explicit cuts and emits both speakers in one token sequence, but the
acoustic encoder is typically not conditioned on speaker activity.
SA-DiCoW~\cite{kocour2026sadicoW} injects a
silence/target/non-target/overlap (STNO) mask, yet the cue remains a hard
frame-level map.
SD-CTC~\cite{sakuma2025sdctc} and speaker-aware CTC~\cite{kang2025sactc}
likewise encourage a hard speaker assignment along time.
A wrong hard assignment mixes two references into one stream or drops words
that were spoken concurrently.

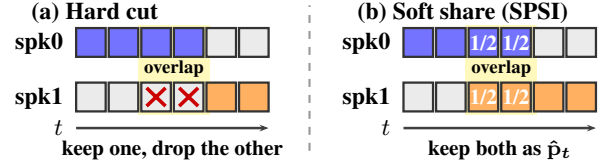
\begin{figure}[t]
\centering
\resizebox{\columnwidth}{!}{\begin{tikzpicture}[
  x=1cm, y=1cm,
  font=\footnotesize\bfseries,
  cell/.style={draw=black!80, line width=0.85pt, inner sep=0pt},
]
\path[use as bounding box] (0,0.02) rectangle (8.50,1.96);

\node[anchor=west, font=\small\bfseries] at (0.02,1.86) {(a) Hard cut};
\node[font=\small\bfseries, text=black, anchor=east] at (0.66,1.46) {spk0};
\node[font=\small\bfseries, text=black, anchor=east] at (0.66,0.74) {spk1};

\fill[yellow!32] (1.54,0.52) rectangle (2.48,1.70);
\node[font=\scriptsize\bfseries, text=black] at (2.01,1.12) {overlap};

\fill[blue!55] (0.72,1.28) rectangle ++(0.39,0.38);
\fill[blue!55] (1.15,1.28) rectangle ++(0.39,0.38);
\fill[blue!55] (1.58,1.28) rectangle ++(0.39,0.38);
\fill[blue!55] (2.01,1.28) rectangle ++(0.39,0.38);
\fill[black!8] (2.44,1.28) rectangle ++(0.39,0.38);
\fill[black!8] (2.87,1.28) rectangle ++(0.39,0.38);
\foreach \x in {0.72,1.15,1.58,2.01,2.44,2.87} {
  \draw[cell] (\x,1.28) rectangle ++(0.39,0.38);
}
\fill[black!8] (0.72,0.56) rectangle ++(0.39,0.38);
\fill[black!8] (1.15,0.56) rectangle ++(0.39,0.38);
\fill[black!8] (1.58,0.56) rectangle ++(0.39,0.38);
\fill[black!8] (2.01,0.56) rectangle ++(0.39,0.38);
\fill[orange!62] (2.44,0.56) rectangle ++(0.39,0.38);
\fill[orange!62] (2.87,0.56) rectangle ++(0.39,0.38);
\foreach \x in {0.72,1.15,1.58,2.01,2.44,2.87} {
  \draw[cell] (\x,0.56) rectangle ++(0.39,0.38);
}
\foreach \x in {1.58,2.01} {
  \draw[red!80!black, line width=1.35pt]
    (\x+0.07,0.63) -- (\x+0.32,0.87)
    (\x+0.07,0.87) -- (\x+0.32,0.63);
}
\draw[black!75, line width=0.95pt, -{Stealth[length=3.4pt]}]
  (0.72,0.34) -- (3.26,0.34);
\node[font=\small\bfseries, text=black, anchor=east] at (0.66,0.34) {$t$};
\node[font=\footnotesize\bfseries, text=black] at (1.99,0.05) {keep one, drop the other};

\draw[black!40, dashed, line width=0.8pt] (3.8,0.04) -- (3.8,1.96);

\node[anchor=west, font=\small\bfseries] at (4.32,1.86) {(b) Soft share (SPSI)};
\node[font=\small\bfseries, text=black, anchor=east] at (4.98,1.46) {spk0};
\node[font=\small\bfseries, text=black, anchor=east] at (4.98,0.74) {spk1};
\fill[yellow!32] (5.86,0.52) rectangle (6.80,1.70);
\node[font=\scriptsize\bfseries, text=black] at (6.33,1.12) {overlap};

\fill[blue!55] (5.04,1.28) rectangle ++(0.39,0.38);
\fill[blue!55] (5.47,1.28) rectangle ++(0.39,0.38);
\fill[blue!55] (5.90,1.28) rectangle ++(0.39,0.38);
\fill[blue!55] (6.33,1.28) rectangle ++(0.39,0.38);
\fill[black!8] (6.76,1.28) rectangle ++(0.39,0.38);
\fill[black!8] (7.19,1.28) rectangle ++(0.39,0.38);

\fill[black!8] (5.04,0.56) rectangle ++(0.39,0.38);
\fill[black!8] (5.47,0.56) rectangle ++(0.39,0.38);
\fill[orange!62] (5.90,0.56) rectangle ++(0.39,0.38);
\fill[orange!62] (6.33,0.56) rectangle ++(0.39,0.38);
\fill[orange!62] (6.76,0.56) rectangle ++(0.39,0.38);
\fill[orange!62] (7.19,0.56) rectangle ++(0.39,0.38);

\foreach \x in {5.04,5.47,5.90,6.33,6.76,7.19} {
  \draw[cell] (\x,1.28) rectangle ++(0.39,0.38);
  \draw[cell] (\x,0.56) rectangle ++(0.39,0.38);
}
\node[font=\footnotesize\bfseries, text=white] at (6.095,1.47) {1/2};
\node[font=\footnotesize\bfseries, text=white] at (6.525,1.47) {1/2};
\node[font=\footnotesize\bfseries, text=white] at (6.095,0.75) {1/2};
\node[font=\footnotesize\bfseries, text=white] at (6.525,0.75) {1/2};

\draw[black!75, line width=0.95pt, -{Stealth[length=3.4pt]}]
  (5.04,0.34) -- (7.58,0.34);
\node[font=\small\bfseries, text=black, anchor=east] at (4.98,0.34) {$t$};
\node[font=\footnotesize\bfseries\boldmath, text=black] at (6.31,0.05) {keep both as $\hat{\mathbf{p}}_t$};
\end{tikzpicture}}
\vspace{-8pt}
\caption{Same overlapping mixture, two assignments.
(a)~Hard cut: overlap frames go to one speaker; the other is dropped.
(b)~SPSI: both speakers kept as a share $\hat{\mathbf{p}}_t{\approx}(1/2,1/2)$.}
\label{fig:hardsoft}
\vspace{-8pt}
\end{figure}

The problem we address is illustrated in Fig.~\ref{fig:hardsoft}.
In a hard cut~(a), each overlap frame is given to one speaker and the other
source is dropped.
In a soft share~(b), the same frames keep both sources as a mixture-normalized
posterior $\hat{\mathbf{p}}_t$, and overlap appears as mass near $(1/2,1/2)$.
The goal is to condition a pretrained encoder--decoder on this continuous share,
without a hard cut and without an external diarizer at inference.

We propose Soft Posterior Speaker Injection (SPSI).
A Soft Posterior Head reads an unconditioned encoding of the mixture and predicts
per-frame posteriors $\hat{\mathbf{P}}$, so overlap is a share rather than a
winner-take-all label.
Multi-layer Feature-wise Linear Modulation (MFLM) injects each row
$\hat{\mathbf{p}}_t$ after selected Whisper encoder blocks: an affine transform
conditions mid-level acoustics and late features before they become decoder memory.
Speaker Memory Prompts (SMP) pool $\hat{\mathbf{P}}$ into a short prompt that
informs the SOT decoder of the utterance-level speaker composition.
A two-pass encoding on shared weights first estimates the share and then
re-encodes with MFLM.
The same $\hat{\mathbf{P}}$ also conditions the decoder at token resolution through SMP.

On two-speaker LibriSpeech overlap, SPSI reduces concatenated
minimum-permutation word error rate (cpWER) from $61.5\%$ to $60.0\%$ under
high overlap and from $51.9\%$ to $51.0\%$ on the full set, relative to the
same-backbone SOT.
Speaker CE, SD-CTC, SA-DiCoW, and Pipeline (oracle/est.\ VAD) do not outperform SOT.
Freeze-posterior overlap-heavy adaptation on LibriCSS~\cite{chen2020continuous}
reaches $36.8\%$ held-out cpWER versus $42.3\%$ for SOT.
Ablations show that MFLM and SMP are complementary, and that multi-layer
injection helps most when overlap is heavy.

The contributions of this paper are as follows:
\begin{itemize}
\setlength{\itemsep}{1pt}
\setlength{\parsep}{0pt}
\setlength{\topsep}{2pt}
\item We propose SPSI to predict a soft speaker share $\hat{\mathbf{P}}$ and inject it into Whisper through MFLM and SMP, without an external diarizer.
\item We use a two-pass encoding on shared weights. Pass~$1$ reads an unconditioned mixture encoding and estimates $\hat{\mathbf{P}}$; pass~$2$ re-encodes the same spectrogram with MFLM, so the conditioner and the recognizer stay in one representation space.
\item We show gains over same-backbone SOT, Speaker CE, SD-CTC, and SA-DiCoW. Freeze-posterior overlap-heavy adaptation further yields a $5.5$-point held-out LibriCSS reduction.
\end{itemize}

\section{Proposed Method}
\label{sec:method}
\begin{figure*}[t]
\centering
\resizebox{\textwidth}{!}{%
\begin{tikzpicture}[
  x=1cm, y=1cm,
  font=\scriptsize,
  >=Stealth,
  every node/.style={outer sep=0.7pt},
  bx/.style={draw=black!50, rounded corners=1.4pt, align=center,
             inner sep=1.8pt, line width=0.4pt},
  inpt/.style={bx, fill=black!6, minimum height=0.60cm},
  enc/.style={bx, fill=blue!13, minimum height=0.60cm},
  head/.style={bx, fill=orange!20, minimum height=0.82cm},
  mflm/.style={bx, fill=green!24, font=\scriptsize\bfseries,
               minimum height=0.60cm, inner sep=1.4pt},
  tok/.style={bx, fill=yellow!18, inner sep=2.3pt, minimum height=0.40cm,
              font=\tiny},
  arr/.style={-{Stealth[length=3.2pt]}, thick, draw=black!62},
  parr/.style={-{Stealth[length=2.8pt]}, line width=0.8pt,
               draw=orange!75!black},
]
\path[use as bounding box] (0,-0.16) rectangle (17.56,7.02);

\node[draw=black!25, fill=black!2.2, rounded corners=2.4pt, line width=0.45pt,
      minimum width=17.44cm, minimum height=3.22cm, anchor=south west]
  at (0.06,3.74) {};
\node[anchor=north west, font=\small\bfseries] at (0.22,6.90)
  {(a) Two-pass encoding};

\node[font=\scriptsize\bfseries, text=black!88, anchor=west] at (0.16,6.00) {pass 1};
\node[inpt, minimum height=0.82cm] (mix) at (1.55,6.00) {mixture $x$};
\node[inpt, minimum height=0.82cm] (mel1) at (3.28,6.00) {Mel $\mathbf{M}$};
\node[enc, minimum width=5.15cm, minimum height=0.82cm] (enc1) at (7.28,6.00)
  {Whisper Encoder \\[0.6pt]
   {\tiny stem${+}$24 blocks${+}$post-LN $\to\mathbf{F}\in\mathbb{R}^{T\times d}$}};
\node[head, minimum width=4.72cm] (head) at (12.72,6.00)
  {\textbf{Soft Posterior Head}\\[0.6pt]
   {\tiny LN${\to}$Conv$5$${\to}$GELU${\to}$Conv$3$${\to}$GELU${\to}$Linear $256{\to}2$}};
\node[head, fill=orange!42, minimum width=1.38cm] (phat) at (16.32,6.00)
  {$\hat{\mathbf{P}}$\\[-0.6pt]
   {\tiny softmax}\\[-1.0pt]
   {\tiny $T{\times}2$}};
\draw[arr] (mix) -- (mel1);
\draw[arr] (mel1) -- (enc1);
\draw[arr] (enc1) -- (head);
\draw[arr] (head) -- (phat);

\node[font=\scriptsize\bfseries, text=black!88, anchor=west] at (0.16,4.16) {pass 2};
\node[inpt] (mel2) at (3.28,4.16) {same $\mathbf{M}$};
\node[enc] (stem) at (4.72,4.16) {stem\\[-1pt]{\tiny ${+}$ pos}};
\node[enc, minimum width=1.16cm] (b18) at (6.14,4.16)
  {blocks\\[-1pt]{\tiny $1$--$8$}};
\node[mflm, minimum width=1.38cm] (f8) at (7.68,4.16) {MFLM $\ell{=}8$};
\node[enc, minimum width=1.20cm] (b916) at (9.26,4.16)
  {blocks\\[-1pt]{\tiny $9$--$16$}};
\node[mflm, minimum width=1.48cm] (f16) at (10.90,4.16) {MFLM $\ell{=}16$};
\node[enc, minimum width=1.26cm] (b1724) at (12.56,4.16)
  {blocks\\[-1pt]{\tiny $17$--$24$}};
\node[mflm, minimum width=1.48cm] (f24) at (14.20,4.16) {MFLM $\ell{=}24$};
\node[enc, fill=blue!22, minimum width=1.26cm] (A0) at (15.76,4.16)
  {$\mathbf{A}_0$\\[-1pt]{\tiny post-LN}};
\draw[arr, dashed] (mel1.south) -- (mel2.north);
\draw[arr] (mel2) -- (stem);
\draw[arr] (stem) -- (b18);
\draw[arr] (b18) -- (f8);
\draw[arr] (f8) -- (b916);
\draw[arr] (b916) -- (f16);
\draw[arr] (f16) -- (b1724);
\draw[arr] (b1724) -- (f24);
\draw[arr] (f24) -- (A0);

\draw[line width=0.8pt, draw=orange!75!black] (phat.south) -- (16.32,5.06);
\draw[line width=0.8pt, draw=orange!75!black] (7.68,5.06) -- (16.32,5.06);
\draw[parr] (7.68,5.06) -- (f8.north);
\draw[parr] (10.90,5.06) -- (f16.north);
\draw[parr] (14.20,5.06) -- (f24.north);
\node[font=\tiny, text=orange!70!black, fill=white, inner sep=1.3pt,
      anchor=south]
  at (15.50,5.18) {inject $\hat{\mathbf{P}}$};

\node[draw=black!25, fill=black!2.2, rounded corners=2.4pt, line width=0.45pt,
      minimum width=17.44cm, minimum height=3.72cm, anchor=south west]
  at (0.06,-0.12) {};
\node[anchor=north west, font=\small\bfseries] at (0.22,3.56)
  {(b) MFLM, SMP, and SOT Decoder};

\node[draw=green!50!black!35, fill=green!8, rounded corners=1.5pt,
      inner sep=4.5pt, minimum width=5.38cm, minimum height=2.40cm,
      align=center, anchor=north west]
  (fzoom) at (0.18,3.14) {%
  {\bfseries MFLM (frame $t$, layer $\ell$)}\\[4pt]
  {\scriptsize $\mathbf{g}^{(\ell)}_t=\mathrm{GELU}(\mathbf{W}_1^{(\ell)}\hat{\mathbf{p}}_t+\mathbf{b}_1^{(\ell)})$}\\[2.2pt]
  {\scriptsize $[\boldsymbol{\gamma}^{(\ell)}_t;\boldsymbol{\beta}^{(\ell)}_t]=\mathbf{W}_2^{(\ell)}\mathbf{g}^{(\ell)}_t+\mathbf{b}_2^{(\ell)}$}\\[2.2pt]
  {\scriptsize $\mathbf{H}^{(\ell)}_{t,:}\leftarrow\mathbf{H}^{(\ell)}_{t,:}\odot(1+\tanh\boldsymbol{\gamma}^{(\ell)}_t)+\boldsymbol{\beta}^{(\ell)}_t$}};
\node[draw=purple!50!black!30, fill=purple!8, rounded corners=1.5pt,
      inner sep=4.5pt, minimum width=5.38cm, minimum height=2.40cm,
      align=center, anchor=north west]
  (membox) at (5.86,3.14) {%
  {\bfseries SMP($K{=}4$)}\\[4pt]
  {\scriptsize $\boldsymbol{\mu}{=}\mathrm{mean}_t\hat{\mathbf{p}}_t,\;
         \boldsymbol{\nu}{=}\max_t\hat{\mathbf{p}}_t$}\\[2.2pt]
  {\scriptsize $o{=}\mathrm{mean}_t\hat{p}_{t,0}\hat{p}_{t,1}$}\\[2.2pt]
  {\scriptsize $\mathbf{q}_{1:K}{=}\mathrm{LN}(\mathrm{MLP}([\boldsymbol{\mu};\,\boldsymbol{\nu};\,o]))$}\\[2.2pt]
  {\scriptsize $\mathbf{A}{=}[\mathbf{q}_{1:K};\,\mathbf{A}_0]$}};
\node[draw=blue!50!black!30, fill=blue!8, rounded corners=1.5pt,
      inner sep=4.5pt, minimum width=5.38cm, minimum height=2.40cm,
      align=center, anchor=north west]
  (decbox) at (11.54,3.14) {%
  {\bfseries Decoder cross-attention}\\[4pt]
  {\scriptsize query $\mathbf{u}_i$; keys/values from $\mathbf{A}$}\\[2.2pt]
  {\scriptsize $\mathbf{c}_i{=}\mathrm{softmax}(\mathbf{u}_i\mathbf{A}^{\top}/\sqrt{d})\,\mathbf{A}$}\\[2.2pt]
  {\scriptsize $\mathcal{L}{=}\mathcal{L}_{\mathrm{ASR}}+\lambda\mathcal{L}_{\mathrm{diar}}$}};

\draw[arr] (fzoom.east) -- (membox.west);
\draw[arr] (membox.east) -- (decbox.west);

\draw[arr, dashed, draw=blue!55!black]
  (A0.east) -- (17.20,4.16) -- (17.20,2.18) -- (decbox.east);

\node[tok] (t0) at (1.35,0.30) {prefix};
\node[tok, fill=green!25] (t1) at (4.55,0.30) {\texttt{<spk0>}};
\node[tok] (t2) at (7.45,0.30) {$y^{(0)}_{1:L_0}$};
\node[tok, fill=orange!30] (t3) at (10.35,0.30) {\texttt{<spk1>}};
\node[tok] (t4) at (13.25,0.30) {$y^{(1)}_{1:L_1}$};
\node[tok] (t5) at (15.85,0.30) {\texttt{<eot>}};
\foreach \a/\b in {t0/t1, t1/t2, t2/t3, t3/t4, t4/t5} {
  \draw[-{Stealth[length=2.4pt]}, line width=0.45pt, draw=black!62,
        shorten >=3.0pt, shorten <=3.0pt]
    (\a.east) -- (\b.west);
}
\node[font=\scriptsize\bfseries, text=black!88, fill=black!2.2, inner sep=1.1pt, anchor=west]
  at (16.52,0.30) {SOT $\mathbf{y}$};
\end{tikzpicture}%
}
\caption{Overview of Soft Posterior Speaker Injection (SPSI).
(a)~Two-pass encoding on shared weights: pass~$1$ encodes $\mathbf{M}$ without Multi-layer Feature-wise Linear Modulation (MFLM) and the Soft Posterior Head predicts $\hat{\mathbf{P}}$; pass~$2$ re-encodes the same $\mathbf{M}$, injecting $\hat{\mathbf{P}}$ with MFLM after blocks $\ell{\in}\{8,16,24\}$ to form memory $\mathbf{A}_0$.
(b)~Speaker Memory Prompts (SMP) pool $\hat{\mathbf{P}}$ into $K{=}4$ tokens $\mathbf{q}_{1:K}=\mathrm{LN}(\mathrm{MLP}([\boldsymbol{\mu};\,\boldsymbol{\nu};\,o]))$, concatenate them with $\mathbf{A}_0$ as $\mathbf{A}=[\mathbf{q}_{1:K};\,\mathbf{A}_0]$, and let the SOT decoder query $\mathbf{u}_i$ cross-attend to $\mathbf{A}$.}
\label{fig:arch}
\end{figure*}
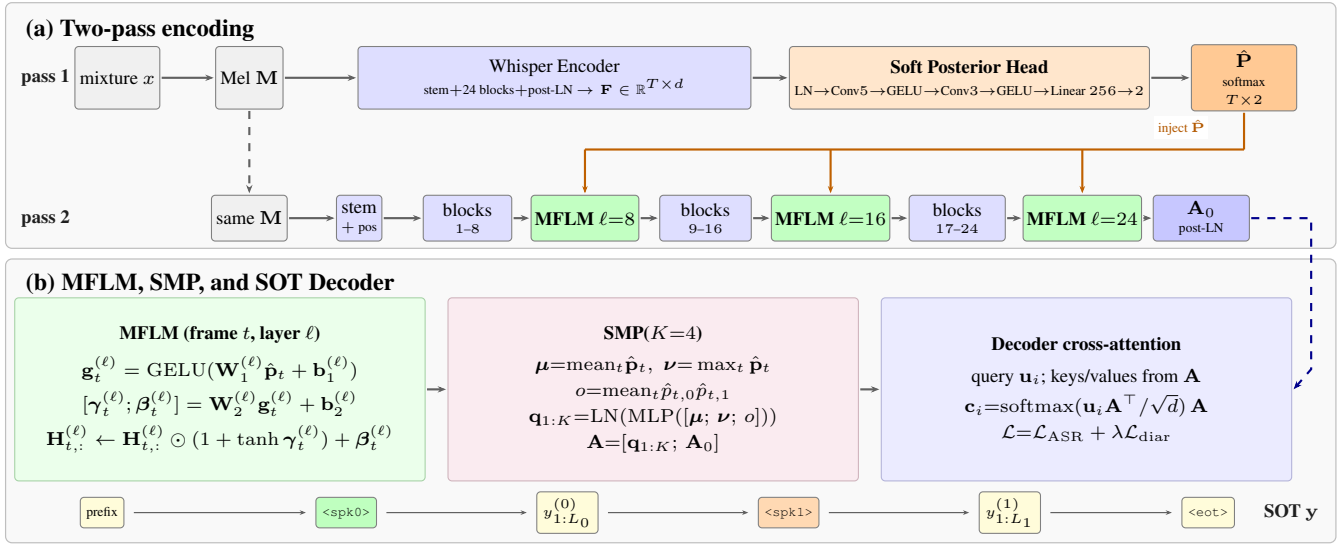

Soft Posterior Speaker Injection (SPSI) comprises three modules, as illustrated in Fig.~\ref{fig:arch}.
The Soft Posterior Head is a lightweight network that maps unconditioned encoder states to a frame-level speaker share $\hat{\mathbf{P}}$.
Multi-layer Feature-wise Linear Modulation (MFLM) injects each row $\hat{\mathbf{p}}_t$ into selected encoder blocks.
Speaker Memory Prompts (SMP) provide a pooled summary of $\hat{\mathbf{P}}$ to decoder cross-attention.
The three modules are trained jointly with a combined ASR and diarization objective. At inference the speaker cue is the predicted $\hat{\mathbf{P}}$; no external diarizer is used.

Let $x$ be a $16$\,kHz mixture and $\mathbf{M}=\mathrm{Mel}(x)$ its $80$-bin log-Mel spectrogram. The Whisper encoder has sequence length $T$ and hidden width $d$. Let $y^{(s)}_{1:L_s}$ be the word-piece transcript of speaker $s$ ($L_s$ pieces; order fixed in the mixture construction).
SOT serializes the two speakers into
\begin{equation}
\begin{aligned}
\mathbf{y}=\bigl[&\mathrm{prefix},\,
\texttt{<spk0>},\, y^{(0)}_{1:L_0},\\
&\texttt{<spk1>},\, y^{(1)}_{1:L_1},\, \texttt{<eot>}\bigr],
\end{aligned}
\label{eq:sot}
\end{equation}
where $\mathrm{prefix}$ is Whisper's English transcribe start,
\texttt{<spk0>} and \texttt{<spk1>} are speaker tags, and \texttt{<eot>} ends the sequence. We write $\hat{\mathbf{P}}$ for the predicted speaker posterior, equivalently a mixture-normalized speaker share, and $\hat{\mathbf{p}}_t$ for its $t$-th row ($t$ indexes encoder frames).

\subsection{Soft Posterior Head}
\label{sec:head}
The Soft Posterior Head estimates a per-frame share instead of assigning each frame to a single speaker. A hard cut would assign each frame to one speaker and drop the weaker source. The predicted row $\hat{\mathbf{p}}_t$ keeps both sources, and overlap corresponds to mass near $(1/2,1/2)$. The share is estimated from the mixture itself, so no external diarizer is required. The head reads an unconditioned Whisper encoding $\mathbf{F}=\mathrm{Enc}(\mathbf{M})$, because feeding $\hat{\mathbf{P}}$ back into this encoding would make the estimate circular. A lightweight temporal stack maps $\mathbf{F}$ to two-speaker logits $\mathbf{Z}\in\mathbb{R}^{T\times 2}$. The stack consists of a layer normalization (LN), two $1$-D convolutions with kernel sizes $5$ and $3$ and hidden size $256$, and GELU activations. A softmax over the two logits yields the posterior,
\begin{equation}
\hat{\mathbf{P}}=\mathrm{softmax}(\mathbf{Z})\in[0,1]^{T\times 2}.
\label{eq:posterior}
\end{equation}
Softmax enforces $\sum_s\hat{p}_{t,s}{=}1$, so $\hat{\mathbf{P}}$ is a
mixture-normalized share rather than two independent activity detectors.

\subsection{Multi-layer Feature-wise Linear Modulation}
\label{sec:mflm}
Concatenating a speaker map only at the encoder output leaves the
pretrained blocks unconditioned; those blocks still mix overlapping
sources as if they were a single talker.
Multi-layer Feature-wise Linear Modulation (MFLM) therefore inserts a per-frame affine
transform~\cite{perez2018film} after blocks $\ell{\in}\{8,16,24\}$.
These sites are early-mid, late-mid, and the encoder output, so that both mid-level
acoustics and late linguistic features can be speaker-conditioned before
they become decoder memory. The hidden width $d$ is left unchanged.
Let $\mathbf{H}^{(\ell)}$ be the hidden states after block $\ell$.
A multi-layer perceptron (MLP) maps $\hat{\mathbf{p}}_t$ to a scale $\boldsymbol{\gamma}^{(\ell)}_t$ and a shift
$\boldsymbol{\beta}^{(\ell)}_t$,
\begin{align}
\mathbf{g}^{(\ell)}_t
&= \mathrm{GELU}(\mathbf{W}_1^{(\ell)}\hat{\mathbf{p}}_t+\mathbf{b}_1^{(\ell)}),
\label{eq:mflm_h}\\
[\boldsymbol{\gamma}^{(\ell)}_t;\boldsymbol{\beta}^{(\ell)}_t]
&= \mathbf{W}_2^{(\ell)}\mathbf{g}^{(\ell)}_t+\mathbf{b}_2^{(\ell)},
\label{eq:mflm_gb}\\
\mathbf{H}^{(\ell)}_{t,:}
&\leftarrow
\mathbf{H}^{(\ell)}_{t,:}\odot(1+\tanh\boldsymbol{\gamma}^{(\ell)}_t)
+\boldsymbol{\beta}^{(\ell)}_t.
\label{eq:mflm}
\end{align}
Here $\mathbf{W}_1^{(\ell)}$, $\mathbf{W}_2^{(\ell)}$, $\mathbf{b}_1^{(\ell)}$, and $\mathbf{b}_2^{(\ell)}$ are learnable, and $\odot$ denotes element-wise multiplication.
The scale $\{1+\tanh\boldsymbol{\gamma}^{(\ell)}_t\}$ stays near one when
$\boldsymbol{\gamma}^{(\ell)}_t{\approx}\mathbf{0}$, so pretrained features are not overwritten
at the start of training.

A two-pass encoding is required because $\hat{\mathbf{P}}$ is not known a priori.
Pass~$1$ computes $\hat{\mathbf{P}}$ without MFLM; pass~$2$ re-encodes the
same $\mathbf{M}$ with Eq.~\eqref{eq:mflm}, yielding encoder memory
$\mathbf{A}_0\in\mathbb{R}^{T\times d}$.
The two passes share encoder weights so that the conditioner and the
recognizer stay in the same representation space.

\subsection{Speaker Memory Prompts}
\label{sec:prompts}
The SOT decoder must decide which speaker token to emit next.
We therefore compress $\hat{\mathbf{P}}$ into a short prompt
that every decoder step can read through cross-attention.

We pool $\hat{\mathbf{P}}\in[0,1]^{T\times 2}$ into three utterance-level statistics:
$\boldsymbol{\mu}=\mathrm{mean}_{t}\hat{\mathbf{p}}_t\in\mathbb{R}^{2}$ (average share),
$\boldsymbol{\nu}=\max_{t}\hat{\mathbf{p}}_t\in\mathbb{R}^{2}$ (element-wise peak share),
and $o=\mathrm{mean}_{t}\,\hat{p}_{t,0}\hat{p}_{t,1}\in\mathbb{R}$ (overlap cue).
An MLP with hidden width $d$ maps the concatenated summary $[\boldsymbol{\mu};\boldsymbol{\nu};o]\in\mathbb{R}^{5}$ to $\mathbb{R}^{Kd}$, and reshaping followed by LN yields $K{=}4$ prompt tokens of width $d$.
At decoder step $i$, query $\mathbf{u}_i\in\mathbb{R}^{d}$ cross-attends to the concatenated memory,
\begin{align}
\mathbf{q}_{1:K}
&=
\mathrm{LN}\bigl(\mathrm{MLP}([\boldsymbol{\mu};\,\boldsymbol{\nu};\,o])\bigr)\in\mathbb{R}^{K\times d},
\label{eq:q}\\
\mathbf{A}
&=
[\mathbf{q}_{1:K};\,\mathbf{A}_0]\in\mathbb{R}^{(K+T)\times d},
\label{eq:A}\\
\mathbf{c}_i
&=
\mathrm{softmax}\!\left(\frac{\mathbf{u}_i\mathbf{A}^{\top}}{\sqrt{d}}\right)\mathbf{A}.
\label{eq:xattn}
\end{align}
The context $\mathbf{c}_i$ enters the decoder residual path, so both
speaker-memory tokens and acoustic frames are available when predicting $\mathbf{y}$.

\subsection{Training Objective}
\label{sec:obj}
The head is supervised by an energy-ratio share.
Let $e_{t,s}$ be the mean-square energy of source $s$ in a $20$\,ms frame, linearly interpolated to encoder length $T$, and $\mathcal{T}_{\mathrm{act}}=\{t:e_{t,0}+e_{t,1}{>}0\}$ the active frames.
For $t\in\mathcal{T}_{\mathrm{act}}$,
\begin{equation}
p^\star_{t,s}=e_{t,s}/(e_{t,0}+e_{t,1}).
\label{eq:pstar}
\end{equation}
The diarization loss is a cross-entropy on $\mathcal{T}_{\mathrm{act}}$, so silent frames do not force a speaker choice,
\begin{equation}
\mathcal{L}_{\mathrm{diar}}
=
\frac{1}{|\mathcal{T}_{\mathrm{act}}|}\sum_{t\in\mathcal{T}_{\mathrm{act}}}
\Bigl(-\sum_s p^\star_{t,s}\log\hat{p}_{t,s}\Bigr).
\label{eq:diar}
\end{equation}
With teacher forcing, the decoder consumes $(\mathbf{y}_{<i},\mathbf{A})$ and a softmax over the Whisper vocabulary yields next token probabilities $\hat{\pi}_{i,y_i}$. Let $\mathcal{I}$ be the content positions of $\mathbf{y}$ (transcribe prefix and pads excluded). The automatic speech recognition (ASR) and joint objectives are
\begin{align}
\mathcal{L}_{\mathrm{ASR}}&= -\sum_{i\in\mathcal{I}}\log \hat{\pi}_{i,y_i},
\label{eq:asr}\\
\mathcal{L}&=\mathcal{L}_{\mathrm{ASR}}+\lambda\mathcal{L}_{\mathrm{diar}},\qquad\lambda{=}0.5.
\label{eq:joint}
\end{align}

\section{Experiments}
\label{sec:exp}

\subsection{Data and Metrics}
\label{sec:data}
Two-speaker mixtures are formed from different-speaker utterances in the clean subsets of LibriSpeech~\cite{panayotov2015librispeech}.
Mixing uses $\mathrm{SIR}\sim\mathrm{Unif}[-5,5]$\,dB, $\mathrm{RT}_{60}\sim\mathrm{Unif}[0.25,0.75]$\,s, and noise at $\mathrm{SNR}\sim\mathrm{Unif}[5,18]$\,dB.
Let $\delta_t{=}1$ if both sources are active in a $20$\,ms frame and $0$ otherwise. With $T_f$ such frames,
\begin{equation}
\rho=\frac{1}{T_f}\sum_{t=1}^{T_f}\delta_t,
\label{eq:rho}
\end{equation}
low, mid, and high bins are $\rho{<}0.25$, $\rho\in[0.25,0.45)$, and $\rho{\ge}0.45$.
Train, development, and test contain $12$k, $1$k, and $1$k mixtures.
Development and test are quota-sampled to $25/40/35\%$ low/mid/high, hence $n{=}250$, $400$, and $350$ per bin.
We fine-tune Whisper-medium for $6$ epochs with AdamW.
The reported checkpoint for every learned system is the one with the lowest high-bin development cpWER.

We use cpWER~\cite{watanabe2020chime,vonneumann2025wer} as the primary metric.
On sentence $j$ the two streams are concatenated and scored under speaker permutation $\sigma$. Let $E_{j,\sigma}$ be the summed substitutions, deletions, and insertions, and $R_j$ the reference-word count. Then
\begin{equation}
\mathrm{cpWER}=\frac{1}{n}\sum_{j=1}^{n}\min_{\sigma}\frac{E_{j,\sigma}}{R_j}.
\label{eq:cpwer}
\end{equation}
Here each sentence is scored independently, so Eq.~\eqref{eq:cpwer} is a per-sentence average rather than a corpus-level micro-average.
Let $\Delta_{\mathrm{cp}}=\mathrm{cpWER}_{\mathrm{SOT}}-\mathrm{cpWER}_{\mathrm{SPSI}}$.
We report two-sided $p$-values from a paired bootstrap with $5000$ replicates, testing $H_0{:}\,\Delta_{\mathrm{cp}}{=}0$.

\subsection{Comparisons on Synthetic Overlap}
\label{sec:compare}
Table~\ref{tab:compare} compares SPSI with prior methods on the same overlap data.
Learned systems use the same Whisper-medium backbone and the same epoch budget, so differences arise from the speaker interface rather than from a stronger recognizer.
SOT~\cite{kanda2020serialized} uses $\lambda{=}0$ and no injector.
Speaker CE keeps the SOT decoder and adds a hard, mutually exclusive per-frame speaker classification loss; encoder features are not modulated.
SD-CTC~\cite{sakuma2025sdctc} is a $29$-symbol character CTC of weight $0.3$ on the same encoder. It is not the official Conformer recipe.
SA-DiCoW~\cite{kocour2026sadicoW} uses oracle STNO masks to drive frame-level diarization-dependent transforms, concatenates speaker channels, and decodes with our SOT tags. It is not the official large-v3-turbo system.
Pipeline (oracle VAD) and Pipeline (est.\ VAD) cut the mixture, recognize each stream with pretrained Whisper, and merge into SOT format.

\begin{table}[ht]
\centering
\setlength{\tabcolsep}{3.5pt}
\caption{Comparison on overlapped LibriSpeech, $n{=}1000$, with low/mid/high bins of size $250/400/350$. Lowest cpWER per column is in bold.}
\label{tab:compare}
\vspace{6pt}
\resizebox{0.98\columnwidth}{!}{%
\begin{tabular}{@{}lcccc@{}}
\toprule
Method & All & Low & Mid & High \\
\midrule
SOT~\cite{kanda2020serialized} & $0.519$ & $0.369$ & $0.529$ & $0.615$ \\
Speaker CE & $0.552$ & $0.397$ & $0.578$ & $0.633$ \\
SD-CTC~\cite{sakuma2025sdctc} & $0.538$ & $0.375$ & $0.561$ & $0.628$ \\
SA-DiCoW~\cite{kocour2026sadicoW} & $0.764$ & $0.566$ & $0.789$ & $0.877$ \\
Pipeline (oracle VAD) & $0.736$ & $0.587$ & $0.762$ & $0.811$ \\
Pipeline (est.\ VAD) & $1.018$ & $1.027$ & $1.006$ & $1.026$ \\
\midrule
\textbf{SPSI (ours)} & $\mathbf{0.510}$ & $\mathbf{0.361}$ & $\mathbf{0.524}$ & $\mathbf{0.600}$ \\
\bottomrule
\end{tabular}%
}
\end{table}

\begin{figure}[!b]
\centering
\includegraphics[width=0.72\columnwidth]{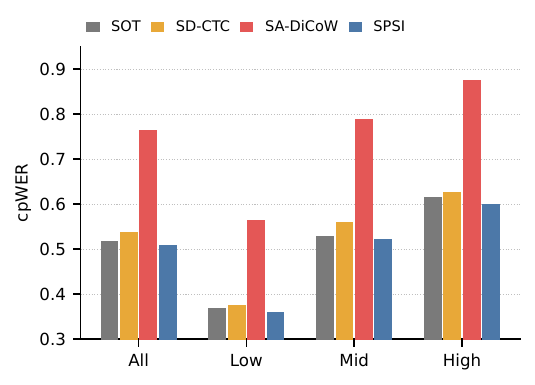}
\vspace{-4pt}
\caption{Synthetic cpWER by overlap bin (Table~\ref{tab:compare}). The SPSI--SOT gap is largest when $\rho{\ge}0.45$.}
\label{fig:cmp_bins}
\vspace{-6pt}
\end{figure}

\begin{figure}[ht]
\centering
\includegraphics[width=0.90\columnwidth]{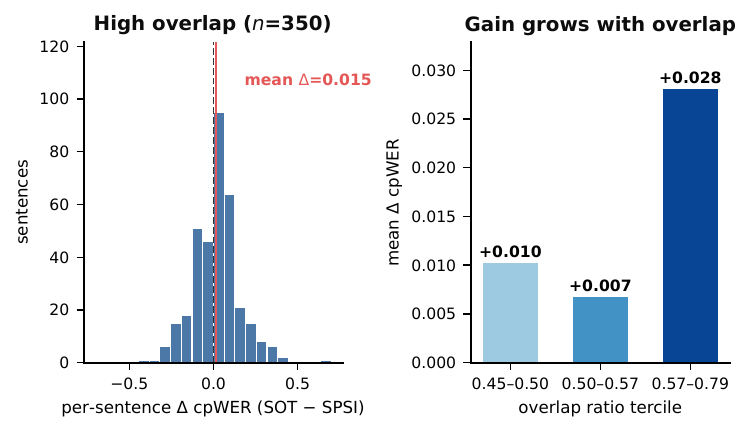}
\vspace{-4pt}
\caption{High-overlap subset ($350$ sentences). Left: per-sentence $\Delta_{\mathrm{cp}}>0$ means SPSI is better. Right: mean $\Delta_{\mathrm{cp}}$ by terciles of $\rho$.}
\label{fig:error}
\vspace{-6pt}
\end{figure}

Table~\ref{tab:compare} and Fig.~\ref{fig:cmp_bins} show that the SPSI advantage is largest under heavy overlap.
Relative to SOT, the high-overlap bin improves by $1.5$ points ($p{=}0.034$) and the full set by $0.9$ points ($p{=}0.029$).
On the $n{=}350$ high-overlap sentences, SPSI is better on $171$, worse on $138$, and tied on $41$ (Fig.~\ref{fig:error}), with the largest mean $\Delta_{\mathrm{cp}}$ in the top tercile of $\rho$.
Speaker CE and SD-CTC increase cpWER in every bin ($0.552$ and $0.538$), both above SOT at $0.519$.
Pipeline (oracle VAD) yields $0.736$ and Pipeline (est.\ VAD) $1.018$: even correct activity still truncates onsets and leaks overlap.
Our SA-DiCoW reimplementation uses oracle STNO at train and test, yet reaches $0.764$, worse than SOT on every bin; because it does not match the official large-v3-turbo system, we read this number as indicative rather than as a faithful measure of that method.
A hard frame assignment is ambiguous when both sources are active; SPSI instead represents overlap as a continuous share, avoiding such ambiguity at inference.

\subsection{Ablation Study}
\label{sec:ablate}
The full model injects $\hat{\mathbf{P}}$ after encoder blocks $\ell{\in}\{8,16,24\}$ and prepends $K{=}4$ SMP tokens.
Table~\ref{tab:ablate} reports ablations that remove one path or reduce the number of MFLM injection sites.

\begin{table}[ht]
\centering
\setlength{\tabcolsep}{3.2pt}
\caption{Ablations of SPSI modules on overlapped LibriSpeech, $n{=}1000$.
Lowest cpWER per column is in bold.}
\label{tab:ablate}
\vspace{6pt}
\begin{tabular*}{0.98\columnwidth}{@{\extracolsep{\fill}}lcccc@{}}
\toprule
Method & All & Low & Mid & High \\
\midrule
\textbf{SPSI} & $\mathbf{0.510}$ & $0.361$ & $0.524$ & $\mathbf{0.600}$ \\
\midrule
Single-layer MFLM & $0.530$ & $0.375$ & $0.538$ & $0.631$ \\
MFLM (no SMP) & $0.531$ & $0.359$ & $0.541$ & $0.642$ \\
SMP (no MFLM) & $0.529$ & $0.380$ & $0.538$ & $0.626$ \\
Single-layer MFLM + SMP & $0.511$ & $\mathbf{0.354}$ & $\mathbf{0.523}$ & $0.611$ \\
\bottomrule
\end{tabular*}
\end{table}
Encoder-only (MFLM no SMP: $0.531$) and decoder-only (SMP no MFLM: $0.529$) variants both underperform the baseline SOT (cpWER $0.519$). SMP (no MFLM) tells the decoder which stream to emit, but the encoder still treats overlap as one talker ($0.529$). MFLM (no SMP) conditions the encoder, yet the SOT decoder has no speaker prompt ($0.531$) and yields the worst high-overlap score of $0.642$. Neither path replaces the other.
Single-layer MFLM without SMP is similarly weak ($0.530$; high $0.631$ versus $0.600$). Single-layer MFLM with SMP recovers most of the average ($0.511$) but remains worse under heavy overlap ($0.611$ versus $0.600$).
A late affine is enough when one speaker dominates a frame, because the pooled SMP prompt can still point the decoder to the right stream.
When both sources occupy the same frames for a long stretch, early and mid encoder blocks have already mixed the two talkers, so a single late affine is not enough. MFLM after blocks $\ell{\in}\{8,16,24\}$ therefore helps most in that regime, and SMP remains necessary to keep the serialized streams aligned. MFLM conditions acoustics at frame resolution, while SMP supplies an utterance-level speaker summary that every decoding step reads through a cross-attention mechanism.

\subsection{Domain Transfer on LibriCSS}
\label{sec:libricss}
We evaluate transfer on real meeting recordings from LibriCSS~\cite{chen2020continuous}.
Monaural recordings are segmented into two-speaker windows of at most $30$\,s with ground-truth time marks.
Sessions $0$--$6$ ($n{=}926$) are used for adaptation, session $7$ for development, and sessions $8$--$9$ ($n{=}259$) as the held-out test in Table~\ref{tab:libricss_adapt_ab}.
Each window is scored as one cpWER sentence.
Let $\phi$ be the Soft Posterior Head and $\theta$ the remaining parameters (Whisper, MFLM, and SMP).
We freeze $\phi$, set $\lambda{=}0$, and adapt $\theta$ on the SOT target $\mathbf{y}$,
\begin{equation}
\min_{\theta}\;
\mathcal{L}_{\mathrm{ASR}}(\mathbf{M},\mathbf{y};\,\hat{\mathbf{P}}_\phi),
\label{eq:freeze}
\end{equation}
so that $\hat{\mathbf{P}}_\phi$ stays calibrated from synthetic pretraining.
Joint fine-tuning instead updates both $\phi$ and $\theta$ with $\lambda{=}0.5$.
SOT has no posterior head; its columns follow the same protocols as their SPSI counterparts (Joint, Freeze, +OV-heavy) with the same data and step budget.
Overlap-heavy continuation keeps $\phi$ frozen and trains further on the OV$20$--OV$40$ subsets of LibriCSS.

\begin{table}[ht]
\centering
\setlength{\tabcolsep}{4pt}
\caption{LibriCSS held-out sessions $8$--$9$, $n{=}259$.
Unadapted uses the synthetic-trained checkpoint with no LibriCSS update.
Freeze and +OV-heavy keep $\phi$ fixed.
Lowest cpWER per column is in bold.}
\label{tab:libricss_adapt_ab}
\vspace{6pt}
\begin{tabular*}{0.98\columnwidth}{@{\extracolsep{\fill}}lcccc@{}}
\toprule
Method & Unadapted & Joint & Freeze & +OV-heavy \\
\midrule
SOT & $0.564$ & $\mathbf{0.453}$ & $0.426$ & $0.423$ \\
\midrule
\textbf{SPSI} & $\mathbf{0.562}$ & $0.502$ & $\mathbf{0.377}$ & $\mathbf{0.368}$ \\
\bottomrule
\end{tabular*}
\end{table}

Table~\ref{tab:libricss_adapt_ab} isolates how the share is treated under domain shift.
Unadapted checkpoints sit near SOT ($0.562$ versus $0.564$).
Jointly updating $\phi$ yields $0.502$, worse than SOT at $0.453$, because $\phi$ overwrites the synthetic-trained share.
Freezing $\phi$ and adapting $\theta$ attains $0.377$ versus $0.426$ for SOT.
Overlap-heavy continuation then reaches $0.368$ versus $0.423$, a $\Delta_{\mathrm{cp}}$ of $5.5$ points.
Keeping the share fixed and adapting the injection path is the transfer protocol that helps.
The same data and step budget apply to SOT, so the $5.5$-point gap comes from how the share is treated, not from extra adaptation.

\section{Conclusion and Future Work}
\label{sec:conclude}
This paper proposes Soft Posterior Speaker Injection (SPSI) for multi-talker speech recognition.
A continuous speaker share $\hat{\mathbf{P}}$ is predicted from the mixture and injected into Whisper through MFLM and SMP, without an external diarizer. 
Under domain transfer this advantage requires keeping the soft posterior frozen; jointly re-adapting it overwrites the synthetic-trained share and underperforms SOT.
Future work includes more than two speakers, streaming recognition, and coupling the Soft Posterior Head to automatic diarization.

\bibliographystyle{IEEEbib}
\newpage
\balance
\bibliography{refs}

\end{document}